\documentclass[aps,prl,twocolumn,floatfix]{revtex4}

\usepackage{amsfonts,amsmath,amssymb,amsthm,amscd}
 \usepackage{acronym,array,bm,color,dcolumn}
 \usepackage{epsfig,graphicx,nomencl,upgreek,url}
\usepackage{subfigure}

\newcommand{\tr}{{\rm tr\thinspace}}
\newcommand{\bra}[1]{\ensuremath{\left\langle{#1}\right\vert}}
\newcommand{\ket}[1]{\ensuremath{\left\vert{#1}\right\rangle}}

\newcommand{\expect}[1]{\ensuremath{\left\langle{#1}\right\rangle}}

\newcommand{\erf}[1]{Eq.~(\ref{#1})}

\newcommand{\beq}{\begin{equation}}
\newcommand{\eeq}{\end{equation}}
\newcommand{\bqa}{\begin{eqnarray}}
\newcommand{\eqa}{\end{eqnarray}}
\newcommand{\nn}{\nonumber}
\newcommand{\eg}{\emph{e.g.},~}
\newcommand{\ie}{\emph{i.e.},~}
\newcommand{\etal}{\emph{et al.}~}

\newcommand{\prsec}[1]{\emph{{#1}} --}

\def\R{\mathbb{R}} 
\def\C{\mathbb{C}}

\def\su{\mathfrak{su}}

\DeclareMathOperator{\imag}{Im}

\newcommand{\ma}[1]{\left[\begin{matrix} #1 \end{matrix}\right]}

\newcommand{\dg}{^{\dagger}}
\newcommand{\trp}{^{\sf T}}

\def\mA{\mathbf{A}}
\def\mx{\mathbf{x}}
\def\mC{\mathbf{C}}
\def\my{\mathbf{y}}
\def\mY{\mathbf{Y}}
\def\mI{\mathbf{I}}
\def\zero{\mathbf{0}}
\def\bnu{\boldsymbol{\nu}}
\def\mH{\mathbf{H}_{rs}}

\newcommand{\SJTU}{Joint Institute of UMich-SJTU and Key Laboratory of
  System Control and Information Processing (MOE),
  Shanghai Jiao Tong University, Shanghai, 200240, China}
\newcommand{\SNL}{Department of Scalable \& Secure Systems Research (08961),
Sandia National Laboratories, Livermore, CA 94550, USA}

\begin{document}

\title{Quantum Hamiltonian identification from measurement time traces}

\author{Jun Zhang$^1$ and
Mohan Sarovar$^2$\footnote{Electronic address:
    \texttt{mnsarov@sandia.gov}}} 
\address{$^1$\SJTU \\ $^2$\SNL}

\date{\today}

\begin{abstract}
\noindent Precise identification of parameters governing quantum
processes is a critical task for quantum information and communication
technologies. In this work we consider a setting where system
evolution is determined by a parameterized Hamiltonian, and the task
is to estimate these parameters from temporal records of a restricted
set of system observables (time traces). Based on the notion of system
realization from linear systems theory we develop a constructive
algorithm that provides estimates of the unknown parameters directly
from these time traces. We illustrate the algorithm and its robustness to measurement noise by applying it to
a one-dimensional spin chain model with variable couplings.
\end{abstract}

\maketitle

\noindent 
The promise of quantum technologies for tasks such as computation,
communication, and metrology is motivating the construction of devices
that are precisely engineered at the nanoscale, and whose quantum
dynamics are exceptionally well characterized and controlled
\cite{mikeandike}. The fragility and sensitivity of typical quantum
devices make achieving such objectives extremely challenging, and
significant research efforts over the past two decades have focused on
addressing these challenges.
 
\begin{figure}
\includegraphics[width=0.9\hsize]{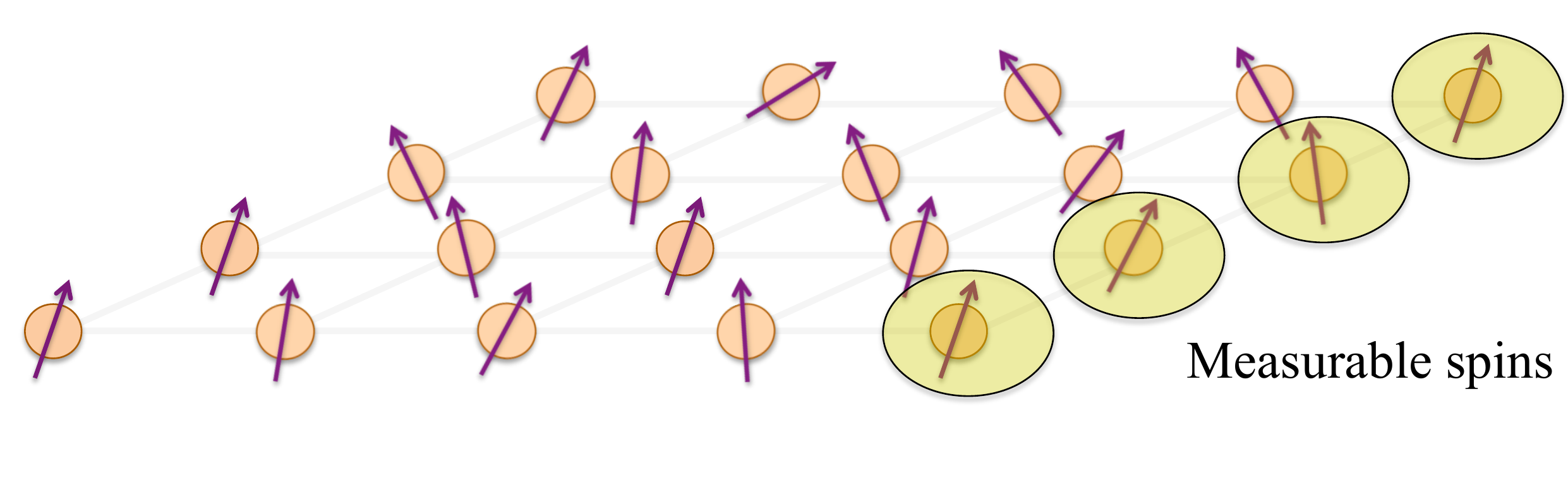}
\caption{\small  A spin (or qubit) lattice as an example illustrating the
  type of system considered in this work. The spins interact with each
  other through nearest-neighbor or long-range couplings and certain
  local observables are measurable for a subset of the spins (circled
  above). The task is to identify the parameters defining the
  Hamiltonian of the interconnected system from a time trace of
  expectation values of these observables.}
\label{fig:spins}
\end{figure}

Process tomography is the most generally applied technique for
characterizing an unknown quantum dynamical process \cite{mikeandike,
  Mohseni:2008ku}.  However, all variants of process tomography are
very resource demanding, \eg in the required number of measurements
settings and number of input state preparations. In addition, it is
often unsuitable in resource-constrained situations where one may only
have measurement access to certain observables or sub-systems; \eg see
Fig.~\ref{fig:spins}.  Furthermore, process tomography does not
utilize often available partial information about the system.  One
such common scenario is when the structure of a dynamical model can be
obtained from underlying physics and what is to be determined are some
unknown parameters in the model. This is the quantum version of
parameter estimation in classical system sciences,
and some previous work has considered variants to quantum tomography
for this problem \cite{daSilva:2011ej}.

In this work, we consider a new approach to quantum parameter
estimation. Whereas process tomography typically measures a complete
basis of system observables at \emph{one} time instant, we ask what
can be achieved if a temporal record of a small set of system
observables is collected? We refer to such a successive record of
observable expectations as an \emph{measurement time trace}, and
develop a method that enables information about dynamical parameters
to be extracted from such time traces. Our method takes into account
{\it a priori} information and fits naturally into resource
constrained situations, and as such we expect that it will be very
experimentally relevant and feasible. Additionally, because our scheme
  utilizes a time trace, it can identify the generator of dynamics
  (\eg a Hamiltonian) as opposed to the dynamical map (\eg a
  unitary at a fixed time), which is typically
  what process tomography achieves. This is advantageous since in
physically realistic scenarios the generator of dynamics is more
compactly specified than the map. This will be discussed in
more detail below.

Several authors have considered parameter estimation from various
types of time-dependent measurement
records~\cite{Boulant:2003ih,Col.Sch.etal-2005,Devitt:06,You.Sar.etal-2009,Burgarth:2009ee,
  Burgarth:2009dl,DiFranco:2009ju,Granade:2012kj,Kato:2013ub,Dominy:2013tk,Jagadish:2014ua}.
Particularly relevant to this work, Cole \etal used Fourier analysis
to identify a single qubit Hamiltonian from one measurement
observable~\cite{Col.Sch.etal-2005}, and Devitt \etal presented a
scheme to identify any two-qubit Hamiltonian from the temporal
evolution of the concurrence measure of entanglement~\cite{Devitt:06}.
Subsequent work by Burgarth \etal \cite{Burgarth:2009ee,
  Burgarth:2009dl} and Di Franco \etal \cite{DiFranco:2009ju}
generalized this approach to estimate the coupling strengths in a
many-qubit network from measurements on a small part of the network.
Recently, Burgarth \etal presented a framework for quantum system
identification based on input/output information and formalized the
notion of equivalence between system
realizations~\cite{Burgarth:2012in}.

Our approach differs from existing work on quantum system
identification in two critical aspects. Firstly, we develop a
constructive algorithm for identification of arbitrary Hamiltonian
quantum dynamics that takes advantage of available prior knowledge of
the system (\eg network structure or partial knowledge of parameters).
The technique can also be employed when such prior information is
absent.  Secondly, in contrast to most existing system identification
schemes, we do not require state tomography of a restricted set of
subsystems, but rather develop a technique that produces parameter
estimates based only on the collected measurement time traces.
  
\prsec{Setup} We consider the task of identifying the Hamiltonian of
an unknown quantum dynamical process. Assume that the dimension of the
system is finite and known, and that the dynamical process can be prepared
at some well-characterized initial states. Further, we assume that the
dynamical evolution of the process is unitary (no decoherence).  This
condition can be relaxed and the approach will be extended to the
non-unitary case in a future publication.

A parameterized form of the Hamiltonian governing the quantum dynamical
process can be written as, 
\begin{equation}
\label{eq:ham_theta}
H= \sum\nolimits_{m=1}^M a_m(\theta) X_m,
\end{equation}
where $\theta$ is a vector consisting of unknown parameters, $a_m \in
\mathbb{R}$ are some known functions of $\theta$, and $X_m$ are known
Hermitian operators \footnote{We set $\hbar=1$ and therefore the $a_m$
  have units $1/s$.}.  Assume that the dimension of the quantum
process is $N$, and thus $iH\in \su(N)$, {\it i.e.} the Lie algebra
consisting of all the $N\times N$ skew-Hermitian matrices.  An
orthonormal basis of $N^2-1$ matrices $\{iX_m\}$ can be chosen for
$\su(N)$, where the Hilbert-Schmidt inner product is defined as
$\langle iX_m, iX_n \rangle \equiv \tr(X_m\dg X_n)$, and hence
$a_m=\tr(H X_m)$. For example, $\frac{i}{2}\sigma_{\alpha}^1\otimes
\sigma_{\beta}^2$ form a basis for the two-qubit algebra $\su(4)$,
where $\sigma_\alpha$, $\sigma_\beta$ can be Pauli matrices
$\sigma_x$, $\sigma_y$, $\sigma_z$, or the identity matrix $I_2$, and
superscripts label the qubits \footnote{In the following we will omit
  the tensor product when writing multi-qubit Pauli operators for
  brevity.}. The numbers $C_{jkl}$ such that
\begin{equation*}
  \label{eq:struct_const}
    [iX_j, iX_k]=\sum\nolimits_{l=1}^{N^2-1} C_{jkl} (iX_l),
\quad j, k=1, \cdots, N^2-1,
\end{equation*}
are the \emph {structure constants} of the Lie algebra $\su(N)$ with
respect to this basis. Each element $X_m$ is Hermitian and thus can be
considered an \textit{observable} for the system. Furthermore, we can
consider the $a_m$ as our unknown parameters, because solving for
$\theta$ from $a_m$ is simply an algebraic problem. 

Note that in Eq.~\eqref{eq:ham_theta}, typically $M \ll N^2 -1$
because of physical constraints on system energy, locality, and weight
of interactions. For instance, the Hamiltonian for the spin lattice
system in Fig.~\ref{fig:spins} contains only weight-one and weight-two
basis elements $X_m$ \footnote{The \textit{weight} of a multi-qubit
  Pauli operator is the number of non-identity terms in the tensor
  product.}, and furthermore, the weight-two interactions might be
restricted to only being between nearest-neighbor spins on the
lattice.  By utilizing measurement time traces our identification
algorithm can estimate the process at the Hamiltonian level where
there are only $M$ unknown parameters. In contrast, process tomography
generally does not consider time traces and therefore must estimate
the process at the unitary level where there are in general $N^2-1$
unknown parameters.

\prsec{Observable dynamics} The dynamics of the expectation value of
an observable $X_k$, written as $x_k=\bra{\psi} X_k \ket{\psi}$, can
be derived as
\begin{equation}
  \label{eq:xk_dyn}
\dot{x}_k
=\sum\nolimits_{l=1}^{N^2-1} 
\left(\sum\nolimits_{m=1}^{M}  C_{mkl} a_m \right)x_l.
\end{equation}
Collecting the $x_k$ in a vector $\mx \in \R^{N^2-1}$, we obtain a
linear equation describing the complete dynamics:
\begin{equation}
  \label{eq:2}
\dot{\mx}= \mA \mx, 
\quad x_k(0)=\bra{\psi(0)} X_k \ket{\psi(0)},
\end{equation}
where the matrix $\mA \in \R^{(N^2-1) \times (N^2-1)}$ has elements
$\mathbf{A}_{kl} = \sum\nolimits_{m=1}^M C_{mkl}a_m$. Using the
antisymmetries of the structure constants, it can be shown that $\mA\trp =
-\mA$.  The vector $\mx$, often called the \emph{coherence
  vector}~\cite{Len-1987}, is a complete representation of the
quantum state . \erf{eq:2} explicitly describes the quantum dynamics as a linear time
invariant (LTI) system and hence it enables application of results from classical linear systems
theory.

Typically, some observable expectation values may be easily measured,
\eg local observables of a collection of spins are tracked as function
of time, see Fig.~\ref{fig:spins}. Often the measured observables
belong to the chosen $\su(N)$ basis, but if not, each observable $O_i$
can be expanded in this basis as $O_i = \sum_j o^{(i)}_j X_j$. Collect
the unique basis elements present in the expansion of all measured
observables in the set $\mathcal{M} = \{X_{\bnu_1}, X_{\bnu_2}, ...,
X_{\bnu_p}\}$, where $\bnu$ is a vector of length $p$. For example, if
$O_1=o^{(1)}_3 X_3 + o^{(1)}_5 X_5$ and $O_2=o^{(2)}_2 X_2 + o^{(2)}_3
X_3$, with $o^{(j)}_k \in \mathbb{R}$, then $p=3$ and $\mathcal{M} = \{X_2, X_3, X_5\}$. Generally,
$p\ll N^2-1$.

In the following we will use time traces of the measured observable expectation values
to identify the unknown Hamiltonian parameters. To this end, we first
need to derive the dynamical equation governing the time evolution of these
observables. Parallel to the study of controllability in classical
nonlinear systems theory~\cite{Sastry:1999wd}, we give a constructive
procedure to obtain the closed dynamics for these
observables. For the Hamiltonian in~\erf{eq:ham_theta}, let
$\Delta=\{X_m\}_{m=1}^M$.  Define an iterative procedure as
\begin{equation}
  \label{eq:34}
G_0=\mathcal{M}, \text{ and }
G_{i}=[G_{i-1}, \Delta] ~\cup~ G_{i-1},
\end{equation}
where $[G_{i-1}, \Delta] \equiv \{X_j : \tr(X_j\dg [g, h]) \neq 0,
~\text{where}~ g\in G_{i-1}, h\in\Delta \}$ \footnote{We do not need
  to keep track of multiplicative constants, only the operators
  generated by these commutators.}. In geometric control theory, the
sequence of $G_i$ are referred to as the \textit{filtration}
associated to $\Delta$~\cite{Sastry:1999wd}.  Since $\su(N)$ is
finite, this iteration will saturate at a maximal set $\bar G$ after
finite steps, and we refer to this set as the \emph{accessible set}.
Intuitively, the set $\bar{G}$ contains the elements of the system
that couple to the measured observables.  Then, writing all the $x_k$
with $X_k \in \bar G$ in a vector $\mx_a$ of dimension $K \leq N^2-1$,
the dynamics for this vector is given by
\begin{equation}
\label{eq:dyn_acc}
\dot{\mx}_a= \tilde\mA \mx_a,
\end{equation}
where $\tilde{\mA}$ is a $K\times K$ sub-matrix of $\mA$, \ie only the
elements necessary to describe the evolution of the subset of
observable averages collected in $\mx_a$. 

\prsec{Identification algorithm} A necessary condition for the identifiability of $a_m$ is that it be
present in the matrix $\tilde \mA$, because otherwise it would not participate in the dynamical equation~\eqref{eq:dyn_acc}, and there would be no way to infer its value from examining the observables in $\mathcal{M}$.  In order to estimate these identifiable
parameters we utilize the notion of a \emph{system realization}
constructed from the measurement time traces. In linear systems theory
there are many methods for constructing a realization of a linear
dynamical system based on measurement results \cite{Callier:1991eo},
and in the following we adapt one of these, the \emph{eigenstate
  realization algorithm} (ERA) \cite{Juang:1985kw}, for the purposes
of Hamiltonian parameter estimation.

The estimation setting we consider is the following.  Suppose we have
access to the expectation values of the observables in $\mathcal{M}$
at regular time instants $j\Delta t$ for some sampling period $\Delta
t$ \cite{SuppInfo}. Denote these values as $\{\my(j\Delta t)\}$, and
they may have to be collected from averaging measurements on several
runs of the experiment under the same initial state. Note that
$\my(j\Delta t)$ is the output of the following discretized form of
Eq.~\eqref{eq:dyn_acc}:
\begin{equation}
\label{eq:dyn_disc}
    \mx_a(j+1) = \tilde\mA_d \mx_a(j), \quad
\my(j) = \mC \mx_a(j),
\end{equation}
where for brevity of notation we use $\mx_a(j) \equiv \mx_a(j \Delta
t)$ and $\my(j) \equiv \my(j \Delta t)$, and $\tilde\mA_d=e^{\tilde
  \mA \Delta t}$. The $p\times K$ matrix $\mC$ picks up the entries in
$\mx_a(j)$ that correspond to expectation values of elements of
$\mathcal{M}$.  Also assume that the system is prepared at a fixed,
known initial state $\mx(0)$, and the corresponding initial state for
Eq.~\eqref{eq:dyn_disc} is $\mx_a(0)$. Then these relations can be
solved easily to obtain an explicit form for the outputs: $\my(j) =
\mC \tilde\mA_d^j \mx_a(0)$.  Having access to the time trace
  $\my(j)$, one may try to solve this set of equations directly.
  However, since $\tilde\mA_d$ is a transcendental function of $a_m$,
  determining the parameters this way is usually infeasible. Instead,
  we will utilize ERA and formulate a new relationship so that
  parameter estimation only requires solving polynomial equations.

The first stage of the estimation algorithm is to construct a minimal
realization of the system based on input/output information. This is
achieved by ERA in three steps, as follows.

\textbf{\underline{Step}} 1: \setlength{\leftmargin}{-10pt} Collect
the measured data into an $rp \times s$ matrix (generalized Hankel
matrix) as:
\begin{equation*}
  \label{eq:hankel}
  \begin{aligned}
 & \mH(k)=\\
&\ma{\my(k)&\my(k+t_1)&\cdots&\my(k+t_{s-1}) \\
\my(j_1+k)&\my(j_1+k+t_1)&\cdots&\my(j_1+k+t_{s-1}) \\
\vdots& \vdots &&\vdots \\
\my(j_{r-1}+k)&\my(j_{r-1}+k+t_1)&\cdots&\my(j_{r-1}+k+t_{s-1})}
  \end{aligned}
\end{equation*}
with arbitrary integers $j_i$ ($i=1$, $\cdots$, $r-1$) and $t_l$
($l=1$, $\cdots$, $s-1$). 

\textbf{\underline{Step}} 2: 
Find the singular value decomposition (SVD) of $\mH(0)$ as
\begin{equation}
  \label{eq:hankel_svd}
  \mH(0)=P\ma{\Sigma&0\\ 0 &0}Q\trp
=\ma{P_1&P_2}\ma{\Sigma&0\\ 0 &0}\ma{Q_1\trp\\ Q_2\trp}, \nn
\end{equation}
where $P\in \R^{rp\times rp}$, $Q\in \R^{s\times s}$ are both
orthonormal, and $\Sigma$ is a diagonal matrix with the non-zero
singular values of $\mH(0)$ determined up to numerical accuracy
$\epsilon$, \ie $\Sigma_{ii}>\epsilon$ for all $i\leq n_\Sigma$ where
$n_\Sigma$ is the dimension of $\Sigma$. The matrices $P_1$, $P_2$,
$Q_1$, $Q_2$ are partitions with compatible dimensions.

\textbf{\underline{Step}} 3: Form a realization of the
system~\eqref{eq:dyn_disc} as $\hat \mA_d = \Sigma^{-\frac12} P_1\trp
\mH(1)Q_1\Sigma^{-\frac12}$, $\hat \mC = \mathsf{E}_{p}\trp P_1
\Sigma^{\frac12}$, where $\mathsf{E}_p\trp = \left[ \mathbf{I}_p, 0_p,
  \cdots, 0_p\right]$.  The pair $(\hat \mA_d, \hat \mC)$ 
reproduces the input-output relations specified by \erf{eq:dyn_disc}, that is:
\begin{eqnarray} 
\my(j) = \mC \tilde \mA_d^j
\mx_a(0) = \hat \mC \hat \mA_d^j \hat \mx(0), \quad \text{for all } j\ge 0,
\label{eq:realization}
\end{eqnarray} 
provided that $\hat \mx(0) \equiv \Sigma^{\frac12} Q_1\trp e_1$, where
$e_1$ is the first column of $\mathbf{I}_{s}$. 

This completes the specification of the ERA algorithm. 
Then let $\hat \mA = \log \hat \mA_d/\Delta
t$~\cite{SuppInfo}.  This results in a realization of the
continuous-time linear
system in the form of the triple $(\hat \mA, \hat \mC, \hat \mx(0))$.
Now, to estimate the Hamiltonian parameters we use an invariant of
different realizations, the \emph{transfer function} \cite{Callier:1991eo}, to form equations for the unknown parameters. Specifically, the transfer function from an
initial state $\mx(0)$ to the measurement observables specified by
$\mC$ can be written as $G(s) = \mC(s\mI - \mA)^{-1}\mx(0)$, where
$s\in \mathbb{C}$ is the Laplace variable. Equating the
transfer functions for the original system with unknown parameters and
the ERA realization we get:
\begin{equation}
  \label{eq:realization_gen}
\mC (s \mI-\tilde \mA)^{-1}\mx_a(0) 
=\hat \mC (s \mI-\hat \mA)^{-1}\hat\mx(0).
\end{equation}
The right hand side of \erf{eq:realization_gen} is completely
determined by the measured data, and the left hand side can be
simplified as the ratio $Q(s)/P(s)$~\cite{Callier:1991eo}, where
\begin{equation}
  \label{eq:1}
 P(s)=\det(s \mI-\tilde\mA), 
Q(s)=\det \left( s\ma{\mI& \zero\\
\zero&\zero} - \ma{\tilde\mA& \mx_a(0)\\
    \mC&\zero}\right). 
\end{equation}
The coefficients of $Q(s)$, $P(s)$ are all polynomials of the
Hamiltonian parameters $a_m$. Equating these coefficients with those
in the right hand side of Eq.~\eqref{eq:realization_gen}, we obtain a system of
polynomial equations. Solving these multivariate polynomial equations leads to the
identification of $a_m$.  
 
A judicious choice for the initial state is crucial to this
identification scheme. For instance, if $\mx_a$ is zero or an
eigenvector of $\tilde \mA$, it leads to no sensitivity in the output
to any of the unknown parameters.  Care must be taken to avoid such
degenerate cases. In fact, running the algorithm with multiple initial
states leads to more polynomial equations with low order and thus
helps to solve these equations more efficiently.

This system identification algorithm can result in multiple estimates
of the unknown parameters, all of which satisfy the input/output
relations captured by \erf{eq:realization_gen}. This is because
several system Hamiltonians can generate the same map between an input
state and measurement time trace, and hence are equivalent from an
input/output perspective \cite{Burgarth:2012in}. When the algorithm
results in multiple parameter estimates and more specification is
needed, one has to appeal to prior information, or add resources such
as additional input states or observable time traces.

\prsec{Example} Consider the following Hamiltonian for a
one-dimensional chain of $n$ qubits:
\begin{equation}
  H=\sum\nolimits_{k=1}^n \frac{\omega_k}{2} \sigma_z^k
+\sum\nolimits_{k=1}^{n-1} \delta_k
  \left(\sigma_+^k\sigma_-^{k+1} + \sigma_-^k \sigma_+^{k+1} \right). \nn
\end{equation}
This Hamiltonian is often used as a model for a spin ``wire'' that
enables quantum state transfer \cite{Bose:2007ji}. Suppose that only
one end of the spin chain is observable, and choose
$\expect{\sigma_x^1}$ as the observable that is tracked. Choosing the
generalized Pauli operators as our basis and calculating the
filtration per \erf{eq:34} yields the accessible set as $\bar{G} = \{
2^{-n/2}\sigma_x^1, 2^{-n/2}\sigma_y^1 \} \cup \{2^{-n/2}\sigma_z^1
\cdots \sigma_z^{k-1}\sigma_x^k, 2^{-n/2}\sigma_z^1 \cdots
\sigma_z^{k-1}\sigma_y^k \}_{k=2}^n$. The system matrix $\tilde \mA$
is $2n\times 2n$ and has the following simple structure
\begin{equation}
  \label{eq:4}
\tilde \mA=\left[\begin{array}{cccccccc}
0&  \omega_1&   0& -\delta_1&   &   &   &   \\
-\omega_1&   0&  \delta_1&   0&  0 &   &   &   \\
0& -\delta_1&   0&  \omega_2&   0& \ddots&   &   \\
\delta_1&   0& -\omega_2&   0&  \ddots&  \ddots&   0&   \\
&   0&   \ddots& \ddots&   \ddots&  \ddots& \ddots& -\delta_{n-1}\\
&   &  \ddots&   \ddots& \ddots&   0&  \delta_{n-1}&   0\\
&   &   &   0&   0& -\delta_{n-1}&   0&  \omega_n\\
&   &   &   &  \delta_{n-1}&   0& -\omega_n&   0
 \end{array}\right] \nn
\end{equation}
with $\mx_a = \left[ \bar x_1, \bar y_1, ..., \bar x_n, \bar y_n
\right]$, where $\bar x_1 \equiv \expect{\sigma_x^1}, \bar
y_1 \equiv \expect{\sigma_y^1}$ and $\bar x_k \equiv \expect{\sigma_z^1
  \cdots \sigma_z^{k-1}\sigma_x^k}, \bar y_k \equiv \expect{\sigma_z^1
  \cdots \sigma_z^{k-1}\sigma_y^k}$ for $k\ge 2$. In this basis $\mC =
\left[ 1, 0, 0, ..., 0 \right]$.  All
parameters in the Hamiltonian appear in $\tilde \mA$, and therefore
the necessary condition for identifying all parameters is satisfied
for an estimation strategy that uses only time traces of
$\expect{\sigma_x^1}$. 

Choosing an initial state
$\frac{|0\rangle+i|1\rangle}{\sqrt{2}}|0\cdots 0\rangle$ (with corresponding coherence vector $[0, 1, 0, \cdots, 0]\trp$), and running ERA
results in a realization $(\hat \mA, \hat \mC,
\hat \mx(0))$. The transfer function is given by
\begin{equation*}
  \label{eq:5}
\mC (s I-\tilde \mA)^{-1}
\mx_a(0)=\frac{q_{2n-2}s^{2n-2}+\cdots+q_2s^2+q_0}
{s^{2n}+p_{2n-2}s^{2n-2}+\cdots+p_2s^2+p_0},   
\end{equation*}
where the detailed expressions of the coefficients $p_i$ and $q_i$ as
polynomials of $\omega_k$ and $\delta_k$ can be calculated via
Eq.~\eqref{eq:1}.  These equations can be solved by mature numerical
toolboxes such as PHCpack~\cite{Verschelde:08} to obtain the unknown
parameters $\omega_k$ and $\delta_k$.  In the Supplementary
Material we simulate time traces for this model with $n=3$ and
solve these polynomial equations to explicitly demonstrate the
parameter estimation algorithm \cite{SuppInfo}. In the absence of
measurement noise, the parameters can be perfectly identified up to
sign of $\delta_k$.  The sign ambiguity is because the coupling
strengths only occur to even order in the polynomial equations when
the local observable being measured is $\expect{\sigma_x^1}$.
Additional measurements or prior information are required to determine
the sign.

\begin{figure}
\centering
\subfigure[~Percentage relative error in mean of estimates]{\includegraphics[width=0.49\hsize]{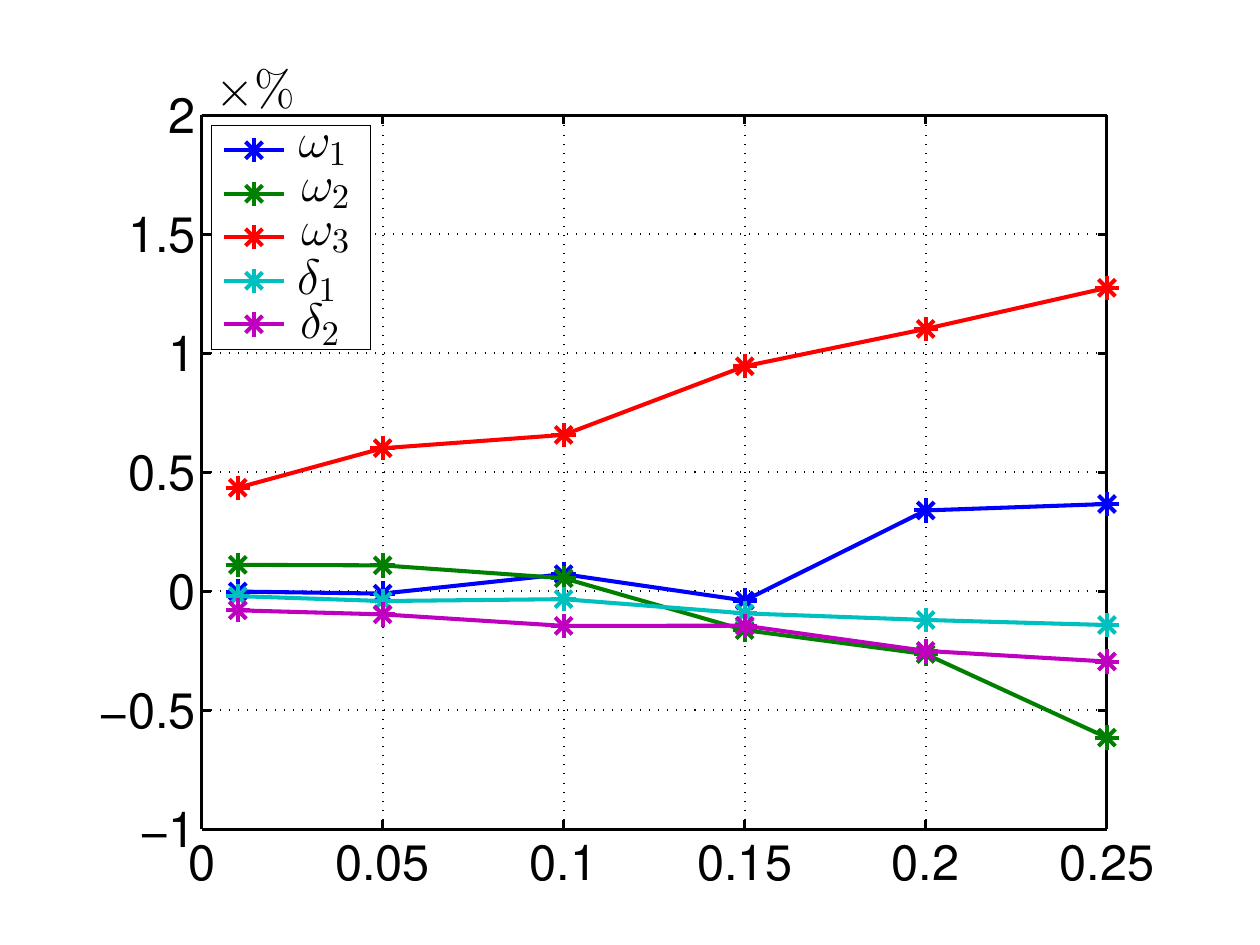}}
\subfigure[~Standard deviation of estimates]{\includegraphics[width=0.49\hsize]{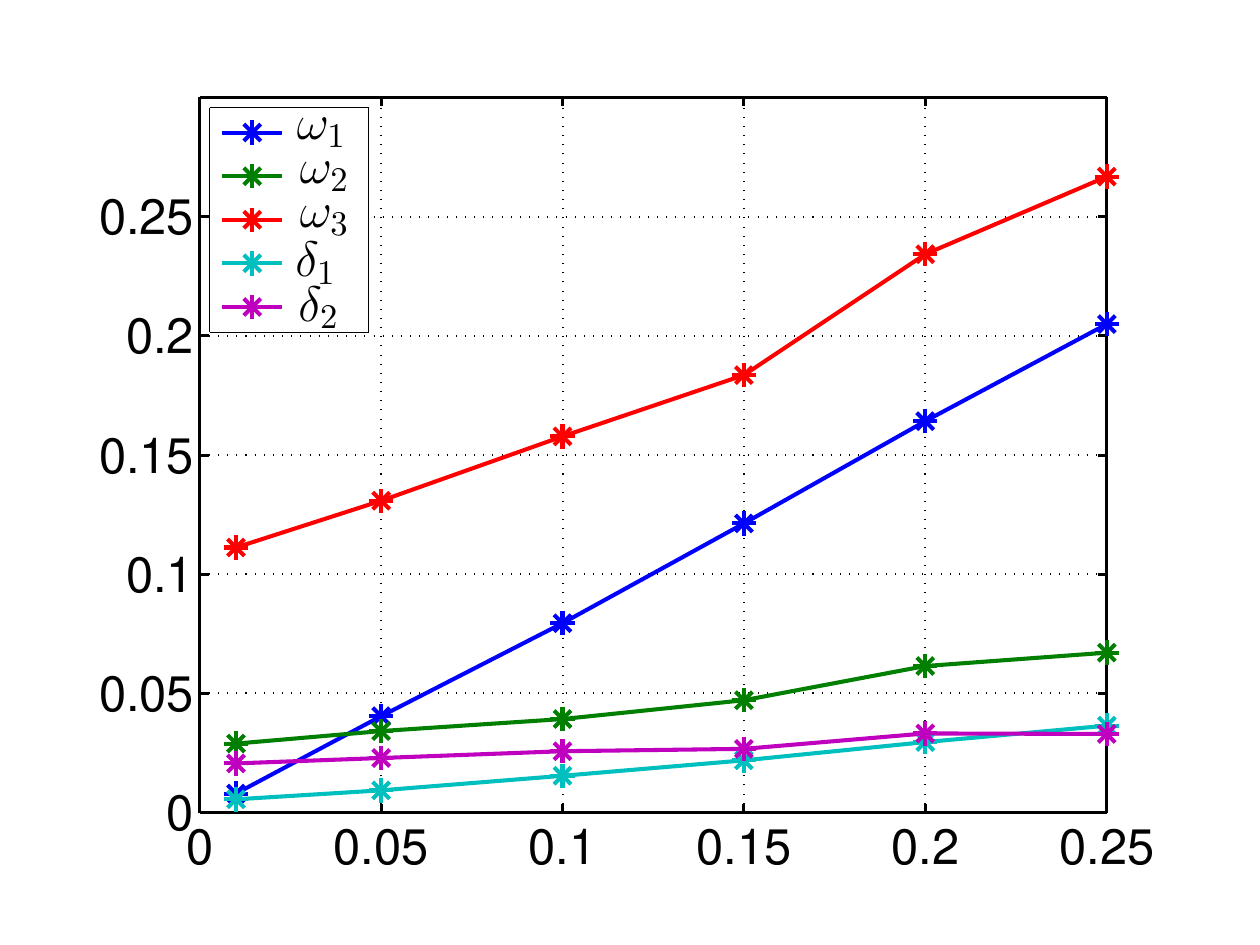}}
\caption{\small Assessing the robustness of parameter estimation algorithm. The $x$-axis
  in both figures is the standard deviation of the measurement noise, $\sigma$.}
\label{fig:noise}
\end{figure}

Experimental measurements of observable expectation values will
inevitably be noisy, and therefore we also assess the performance of
our estimation algorithm in the presence of measurement noise.
Consider the case where the measurements in the $3$-qubit example
specified in the Supplementary Material are corrupted by additive
Gaussian noise, \ie $\my(j)= \expect{\sigma_x^1}(j)+\eta(j)$, with
$\eta(j)\sim \mathcal{N}(0, \sigma)$. The observable
$\expect{\sigma_x^1}(j)$ lies in the range $[-1, 1]$, and we consider
noise with $\sigma$ values 0.01, 0.05, 0.10, 0.15, 0.20, and 0.25.
For each $\sigma$, we generate $4000$ Gaussian noise trajectories and
estimate the five parameters, $\theta = (\omega_1, \omega_2, \omega_3,
\delta_1, \delta_2)$, from each noisy measurement trace.
Fig.~\ref{fig:noise} shows summary statistics that demonstrate the accuracy and robustness of the estimation procedure. The relative error in the mean of the parameter estimates, $\frac{ \bar{ \hat{ \theta_i}} -
\theta_i}{\theta_i}\times 100\%$ \footnote{$\bar{X}$ is the empirical mean of the random variable $X$.}, remains small, whereas the standard deviation of the estimates scales approximately linearly with $\sigma$. Further characterization
of the robustness of the procedure to measurement noise is presented in the Supplementary Material.   
We note that the robustness of our method is a function of the
realization algorithm (ERA) \emph{and} realization invariant used to
construct the polynomial equations. In fact, we experimented with
another invariant, the \emph{Markov parameters} of a system, and discovered
that it is not as robust to noise as the transfer function approach presented
here.

\prsec{Conclusion} We have developed a robust algorithm to identify
the unknown parameters of a quantum Hamiltonian from the time traces
of a set of system observables, which naturally takes into account
prior information and restrictions on measurement access.  A direction
for future work is the generalization of this algorithm to parameter
estimation for open quantum systems governed by Lindblad evolution
\cite{Bre.Pet-2002}, in which case the evolution of the coherence
vector is described by an affine time-invariant system of equations
\cite{Len-1987}.

\prsec{Acknowledgments}
\label{sec:acknowledgements}
MS thanks Akshat Kumar for information on techniques for solving multivariate polynomial systems.
This work was supported by the Laboratory Directed Research and
Development program at Sandia National Laboratories. Sandia is a
multi-program laboratory managed and operated by Sandia Corporation, a
wholly owned subsidiary of Lockheed Martin Corporation, for the United
States Department of Energy's National Nuclear Security Administration
under contract DE-AC04-94AL85000.  JZ acknowledges financial support
from NSFC under Grant No.  61174086, and State Key Laboratory of
Precision Spectroscopy, ECNU, China.  The authors are grateful for
the hospitality of KITP at UCSB, where this work was
initiated. This research was supported in part by the National Science
Foundation under Grant No. NSF PHY11-25915. 
\bibliographystyle{apsrev}
\bibliography{ham_reconstruction}

\pagebreak
\begin{widetext}
\section{Supplementary Information for ``Quantum Hamiltonian identification from measurement time traces"}

\subsection{Choosing the sampling period}
The starting point for our system identification algorithm is a time
trace representing sampled outputs of the system. Forming the
realization $\hat \mA$ is equivalent to reconstructing the continuous
time system, and therefore we expect that a
judicious choice of sampling period $\Delta t$ in the original time
trace is important to obtain accurate results from the algorithm. In
this section we outline the requirements for $\Delta t$.

From $\tilde\mA\trp=-\tilde\mA$, we know that the eigenvalues of
$\tilde\mA$ are all pure imaginary numbers. Therefore the observable
dynamics determined by $\dot \mx_a=\tilde\mA \mx_a$ is a summation of
sinusoidal functions, whose frequencies are given by the eigenvalues
of $\tilde\mA$. To perfectly recover the continuous time dynamics, we
will require the sampling time $\Delta t$ to satisfy the Nyquist
Sampling Theorem \cite{Franklin:1997ua}, which states that the
sampling frequency needs to be at least twice the highest frequency in
the observable dynamics. The highest angular frequency is given by
$\max|\sigma(\tilde\mA)|$, where $\sigma(\tilde\mA)$ denotes the
spectrum of $\tilde\mA$. This in turn yields the corresponding highest
frequency as $\max|\sigma(\tilde\mA)|/2\pi$. Hence, Nyquist Sampling
Theorem imposes a requisite condition on the sampling frequency
$f_{\text{sampling}}$ as \beq
\label{eq:5}
f_{\text{sampling}}>  2 \frac{\max|\sigma(\tilde\mA)|}{2\pi},
\eeq
which leads to
\beq
\label{eq:1}
\Delta t =\frac{1}{f_{\text{sampling}}} <
\frac{\pi}{\max|\sigma(\tilde\mA)|}.  \eeq \erf{eq:1} is a condition
on how to choose the sampling period $\Delta t$, and the right hand
side is a time scale describing the system. Of course, in a
Hamiltonian parameter estimation problem, we usually do not know the
eigenvalues of the matrix $\tilde\mA$. Hence we will need to guess a
suitable sampling time and then refine it with an adaptive method if
necessary.

Choosing a sampling period satisfying \erf{eq:1} becomes
  particular important when taking the matrix logarithm of
  $\hat{\mA}_d$. A sampling period less than required by the Nyquist
Sampling Theorem implies that this logarithm is defined uniquely. To
see this, note that \erf{eq:1} implies \beq
\label{eq:2}
\max|\sigma(\tilde\mA \Delta t)| < \pi.  \eeq Since $(\hat \mA , \hat
\mC, \hat \mx(0))$ from ERA is a minimal realization of the original
system represented by $(\tilde\mA, \mC, \mx_a(0))$, the eigenvalues of
$\hat\mA$ must also be the eigenvalues of $\tilde\mA$. Therefore, we
obtain \beq
\label{eq:8}
\max|\sigma(\hat\mA \Delta t)| < \pi.
\eeq

Now let us quote the following Theorem from Page 20 in Ref.
\cite{Higham:2008ti}, which introduces the notion of principal
logarithm:
\begin{center}\parbox[]{6in}{
    {\bf Theorem 1.31 } Let $A\in \C^{n\times n}$ have no eigenvalues
    on $\R^-$. There is a unique logarithm $X$ of $A$ all of whose
    eigenvalues lie in the strip $\{z: -\pi < \imag(z) < \pi\}$. We
    refer to $X$ as the principal logarithm of $A$ and write
    $X=\log(A)$. If $A$ is real then its principal logarithm is real.}
\end{center}
Here $\R^-$ denotes the negative real axis. Since the eigenvalues of
$\hat\mA$ are purely imaginary, bounded as \erf{eq:8}, and
$\hat{\mA}_d=e^{\hat\mA \Delta t}$, we know that $\hat\mA_d$ has no
eigenvalues on $\R^-$. Therefore, the Theorem above applies and there
is a unique principal logarithm. Furthermore, note that the
conclusion that the eigenvalues of the principal logarithm lie
in the strip $\{z: -\pi < \imag(z) < \pi\}$ is consistent with the
properties of $\hat \mA$ provided that the sampling time is
sufficiently small so as to satisfy \erf{eq:8}.

Therefore we see that the accuracy of the algorithm relies on the
sampling time of the measurement time trace being sufficiently small.

\subsection{Example: three qubit XX spin chain}
In this section we explicitly demonstrate our system identification
algorithm for the spin chain example in the main text, with $n=3$
qubits.

\begin{figure}[tb]
\includegraphics[scale=0.35]{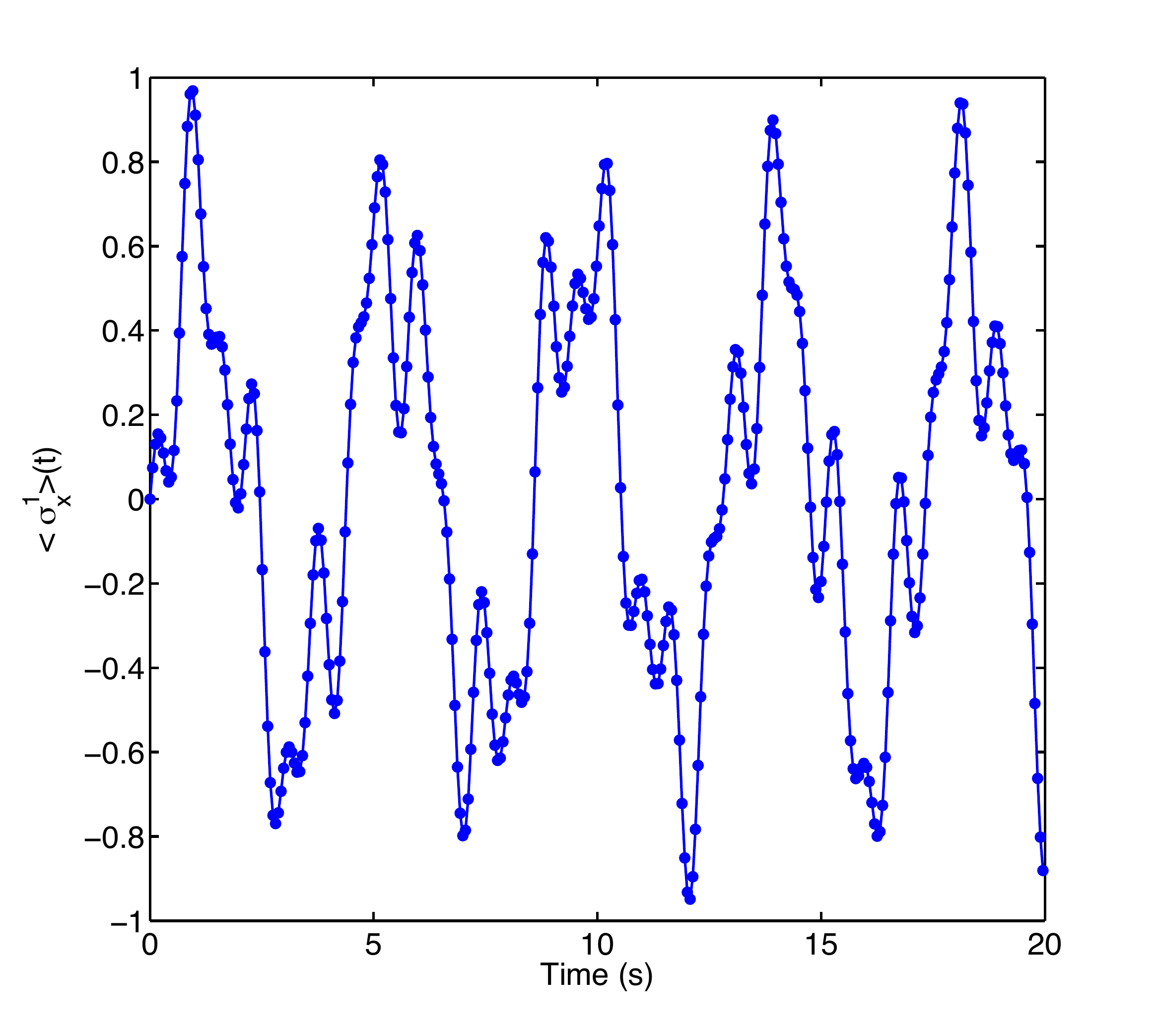}
\caption{\small Measurement time trace for observable 
  $\expect{\sigma^1_x}$ for the XX spin chain example with $n=3$ qubits.
  The dots show a sampled measurement trace for initial state
  $\frac{|0\rangle+i|1\rangle}{\sqrt{2}}|0 0\rangle$.}
\label{fig:time_traces}
\end{figure}

Consider the following Hamiltonian:
\begin{equation}
  H=\sum_{k=1}^3 \frac{\omega_k}{2} \sigma_z^k+\sum_{k=1}^{2} \delta_k
  \left(\sigma_+^k\sigma_-^{k+1} + \sigma_-^k \sigma_+^{k+1} \right),
\end{equation}
with nominal true parameter values $\omega_1 = 1.3$, $\omega_2 = 2.4$,
$\omega_3 = 1.7$, $\delta_1 = 4.3$, $\delta_2 = 5.2$ (all parameters have
units $1/\text{sec}$).

Choose the initial state as
$\frac{|0\rangle+i|1\rangle}{\sqrt{2}}|0\cdots 0\rangle$, and the
corresponding coherence vectors $\mx_a(0)$ is $[0, 1, 0, \cdots,
0]\trp$. Let us assume that we measure the observable $\langle
\sigma_x^1\rangle$ as in the main text. The Laplace transform of the output $\my(t)$ can be written
as 
\begin{equation}
  \label{eq:4}
\mY(s)=\mC (s I-\tilde \mA)^{-1}
\mx_a(0)=\frac{q_4s^4+q_2s^2+q_0}{s^6+p_4s^4+p_2s^2+p_0}, 
\end{equation}
where
\begin{equation}
  \label{eq:6}
  \begin{aligned}
p_4&=2 \delta_1^2+2 \delta_2^2+\omega_1^2+\omega_2^2+\omega_3^2, \\    
p_2&=\delta_1^4+2\delta_1^2\delta_2^2-2 \delta_1^2\omega_1\omega_2 
+2 \delta_1^2 \omega_3^2+\delta_2^4-2 \delta_2^2\omega_2\omega_3 
+2 \delta_2^2 \omega_1^2 
+\omega_1^2\omega_2^2+\omega_2^2\omega_3^2+\omega_1^2\omega_3^2,  \\  
p_0&=\delta_1^4\omega_3^2 +2 \delta_1^2 \delta_2^2\omega_1\omega_3
 -2 \delta_1^2 \omega_1\omega_2 \omega_3^2 +\delta_2^4 \omega_1^2
-2\delta_2^2\omega_1^2\omega_2\omega_3+\omega_1^2\omega_2^2 \omega_3^2, \\  
  \end{aligned}
\end{equation}
and
\begin{equation}
  \label{eq:7}
  \begin{aligned}
q_4&=\omega_1, \\  
q_2&=\omega_1 \omega_2^2-\delta_1^2 \omega_2+2 \delta_2^2 \omega_1+\omega_1\omega_3^2 , \\  
q_0&=- \delta_1^2 \omega_2\omega_3^2+\omega_1\omega_2^2 \omega_3^2 
-2 \delta_2^2 \omega_1\omega_2 \omega_3+\delta_2^4\omega_1
+\delta_1^2 \delta_2^2\omega_3. \\      
  \end{aligned}
\end{equation}

Fig. \ref{fig:time_traces} shows measurement time traces for the
initial state $\frac{|0\rangle+i|1\rangle}{\sqrt{2}}|0 0\rangle$ when
simulated for $T=20s$ with $\Delta t=0.0598 s$.  Using this data we
construct the Hankel matrix $\mH(0)$ with $r=167$, $s=167$ and all
$j_i=1$ and $t_l=1$. Then, performing the remaining ERA steps we
obtain a realization $(\hat \mA_{d}, \hat \mC, \hat \mx(0))$. Further
taking the logarithm results in a realization $(\hat \mA, \hat \mC,
\hat \mx(0))$ of the continuous system. This realization has the same
dimension as the original $\tilde \mA_d$, \ie $n_\Sigma=6$.

For the parameter estimation stage of the algorithm we need to pick
the five lowest order polynomial equations from Eqs.~\eqref{eq:6} and
\eqref{eq:7} (since there are five unknown parameters in this system):
\begin{eqnarray*}
\omega_1 &=&  1.3\\
2 \delta_1^2+2 \delta_2^2+\omega_1^2+\omega_2^2+\omega_3^2 &=& 101.4 \\
\omega_1 \omega_2^2-\delta_1^2 \omega_2+2 \delta_2^2 \omega_1+\omega_1\omega_3^2
&=& 37.173 \\
\delta_1^4+2\delta_1^2\delta_2^2-2 \delta_1^2\omega_1\omega_2 
+2 \delta_1^2 \omega_3^2+\delta_2^4-2 \delta_2^2\omega_2\omega_3 
+2 \delta_2^2 \omega_1^2 
+\omega_1^2\omega_2^2+\omega_2^2\omega_3^2+\omega_1^2\omega_3^2 &=& 1966.4892 \\
- \delta_1^2 \omega_2\omega_3^2+\omega_1\omega_2^2 \omega_3^2 
-2 \delta_2^2 \omega_1\omega_2 \omega_3+\delta_2^4\omega_1
+\delta_1^2 \delta_2^2\omega_3&=& 1407.01176
\end{eqnarray*}
These equations can be solved by mature numerical toolboxes such as \texttt{Singular}, \texttt{Macaulay 2}, \texttt{SOSTools}, and
  \texttt{PHCpack}.
In particular, we applied \texttt{PHCpack} \cite{Verschelde:08} to obtain the following estimates
for the parameters:
\begin{equation*}
  \hat \omega_1 = 1.3, ~\hat \omega_2 = 2.4, 
~\hat \omega_3 = 1.7, ~\hat \delta_1 = \pm 4.3, ~\delta_2 = \pm 5.2.
\end{equation*}
The estimates exactly match the true parameters, except for the
indeterminate sign for the coupling parameters. As discussed in the
main text, this uncertainty in the sign is a result of the equivalence
of systems under some input/output maps, and cannot be resolved unless
additional measurements and/or initial states are introduced.

\subsubsection{Robustness to noise}
\begin{figure}[tb]
\centering
\includegraphics[width=0.6\textwidth]{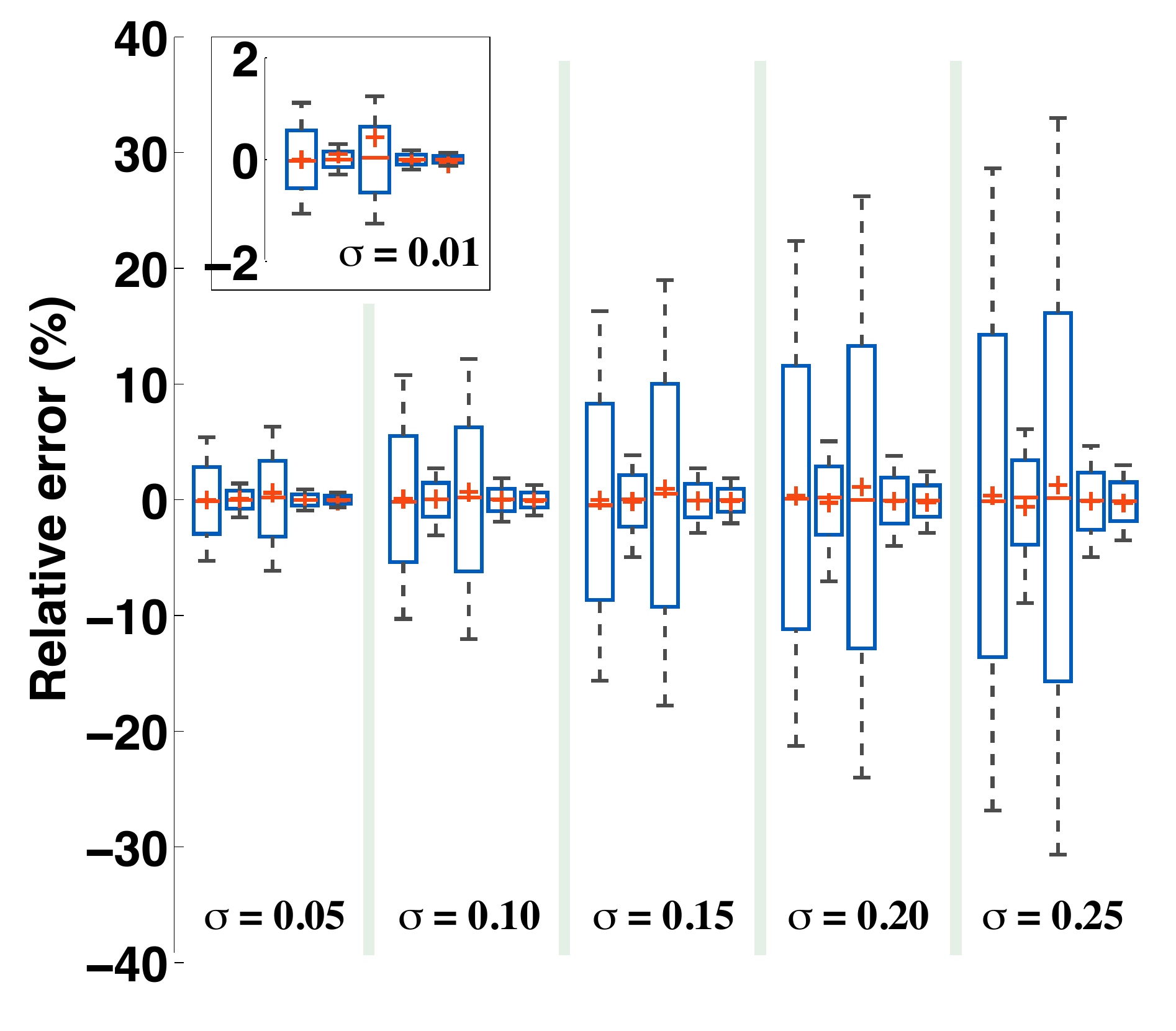}
\caption{Box plots for estimates formed from noisy measurement
  records.  Each group of box plots is for estimation using
  measurement records with noise of standard deviation indicated by
  $\sigma$. The inset shows box plots for $\sigma = 0.01$ separately since the range of relative error in parameter estimates in this case is much smaller than for the other cases. The five box plots in each group are for estimates of parameters (from left to right): $\omega_1, \omega_2, \omega_3,
  \delta_1, \delta_2$.} \label{fig:box_plots}
\end{figure}

As shown in the main text, the Hamiltonian parameter estimation
algorithm we have developed is robust to measurement noise. To
demonstrate this, we perturbed the measurement of observable
$\expect{\sigma_x^1}$ with additive Gaussian noise trajectories; \ie
$\my(j)= \expect{\sigma_x^1}(j)+\eta(j)$, with $\eta(j)\sim
\mathcal{N}(0,\sigma)$. We consider noise with $\sigma$ values 0.01,
0.05, 0.10, 0.15, 0.20, and 0.25.  For each $\sigma$, we generate
$4000$ Gaussian noise trajectories and estimate the five parameters,
$\theta = (\omega_1, \omega_2, \omega_3, \delta_1, \delta_2)$, from
each noisy measurement trace.

Fig. 2 in the main text shows the mean and standard deviation of the
$4000$ estimates for each parameter. To further characterize the
variation of the estimates in that figure, we also shows box plots for
the relative error in estimates, in Fig. \ref{fig:box_plots}. The red $+$ in each box plot indicates the mean
  of the estimates. The red line in each box indicates the median while the bottom and top of each box indicate the $25^{\rm th}$ and $75^{\rm th}$ percentile of the data, respectively. The end points of the whiskers represent the $9^{\rm th}$ and $91^{\rm st}$ percentiles. 
Interestingly, some parameter estimates are
more sensitive to noise than others. In this example, $\hat\omega_1$
and $\hat \omega_3$ are the most sensitive.

\subsection{Note on $n_\Sigma$}
The rank of Hankel matrix, $n_\Sigma$, is the size of the
 reconstructed realization $\hat \mA_d$ (or $\hat \mA$), and is an informative parameter. If
 $n_\Sigma < K$, this means that the original dynamical systems lacks
 complete controllability or observability. An obvious way in which
 this can happen is if, for example, some coupling parameters for a
 network of qubits are actually zero and thus part of the network is
 decoupled from the portion being measured. That part of the system is
 then irrelevant for the dynamics captured in the Hankel matrix and is
 non-identifiable from the measured observables.

\end{widetext}

\end{document}